\documentclass[prl, aps, reprint, superscriptaddress]{revtex4-2}

\usepackage{amsfonts}
\usepackage{amsmath}
\usepackage{mathtools}
\usepackage{float}
\usepackage{color}
\usepackage{gensymb}
\usepackage{multirow}
\usepackage[mathscr]{euscript}
\usepackage{pifont}
\usepackage{mathtools}
\DeclarePairedDelimiter{\ceil}{\lceil}{\rceil}

\newcommand{\ee}[1]{\ensuremath{10^{#1}}}
\newcommand{\dtwomin}[0]{\ensuremath{D^2_{\mbox{min}}}}

\usepackage{empheq}
\usepackage{physics}
\newcommand{\ignore}[1]{}

\begin{document}

\title{Structuro-elasto-plasticity (StEP) model for plasticity in disordered solids}

\author{Ge Zhang}
\affiliation{\emph{Department of Physics}, \emph{City University of Hong Kong}, Hong Kong, China}
\author{Hongyi Xiao}
\affiliation{\emph{Department of Physics and Astronomy}, \emph{University of Pennsylvania},
Philadelphia PA 19104 }
\author{Entao Yang}
\affiliation{\emph{Department of Chemical and Biomolecular Engineering}, \emph{University of Pennsylvania},
Philadelphia PA 19104 }
\author{Robert J. S. Ivancic}
\affiliation{\emph{Materials Science and Engineering Division}, \emph{National Institute of Standards and Technology},
Gaithersburg, MD 20899}
\author{Sean A. Ridout}
\affiliation{\emph{Department of Physics and Astronomy}, \emph{University of Pennsylvania},
Philadelphia PA 19104 }
\author{Robert A. Riggleman}
\affiliation{\emph{Department of Chemical and Biomolecular Engineering}, \emph{University of Pennsylvania},
Philadelphia PA 19104 }
\author{Douglas J. Durian}
\affiliation{\emph{Department of Physics and Astronomy}, \emph{University of Pennsylvania},
Philadelphia PA 19104 }
\author{Andrea J. Liu}
\email{ajliu@physics.upenn.edu}
\affiliation{\emph{Department of Physics and Astronomy}, \emph{University of Pennsylvania},
Philadelphia PA 19104 }

\begin{abstract}
Elastoplastic lattice models for the response of solids to deformation typically incorporate structure only implicitly via a local yield strain that is assigned to each site. However, the local yield strain can change in response to a nearby or even distant plastic event in the system. This interplay is key to understanding phenomena such as avalanches in which one plastic event can trigger another, leading to a cascade of events, but typically is neglected in elastoplastic models. To include the interplay one could calculate the local yield strain for a given particulate system and follow its evolution, but this is expensive and requires knowledge of particle interactions, which is often hard to extract from experiments. Instead, we introduce a structural quantity, ``softness," obtained using machine learning to correlate with imminent plastic rearrangements. We show that softness also correlates with local yield strain. We incorporate softness to construct a "structuro-elasto-plasticity" model that reproduces particle simulation results quantitatively for several observable quantities, confirming that we capture the influence of the interplay of local structure, plasticity, and elasticity on material response.
\end{abstract}

\maketitle

Many disordered solids exhibit ductile behavior when placed under large mechanical load such as shear, and can be strained indefinitely without fracturing. Such systems have stress-strain curves that are monotonic except for fluctuations due to crackling noise arising from avalanches, in which one localized plastic event (particle rearrangement) triggers another and so on. The avalanches are typically described by distributions of energy and stress drops, which have power-law tails reflecting the range of avalanche sizes, from localized to extended across the entire system~\cite{sethna2001crackling,saljeDahmen2014crackling,sethna2017deformation,dahmen2019FrontiersinPhysicsreview}. 

This ductile behavior is often described by a class of models known as elastoplastic (EP) models~\cite{nicolas2017deformation} that 
capture the interplay between elasticity and plasticity: Local regions yield (deform plastically) under sufficient elastic strain, while the plastic deformation generates elastic strain elsewhere. 
EP models incorporate plasticity by imposing a threshold on the amount of strain that a local region can withstand before it yields (the local yield strain). However, such models usually neglect~\cite{martens2012spontaneous}, or highly simplify~\cite{budrikis2013avalanche, martens2021elastoplastic}, the well known effects of both plasticity and elasticity on local yield strain~\cite{barbot2020rejuvenation}. These facilitation effects arise because local yield strain is determined by local structure. Yielding scrambles local structure, so when a region yields, it not only changes its own local yield strain but local yield strains of nearby regions. Yielding also generates elastic strain, which changes the local yield strains of more distant regions. 

A previous study addressed the role of local structure in ductile jammed packings of two-dimensional Hertzian bidisperse disks \cite{zhang2021interplay}. There, the local structure was quantified by softness \cite{schoenholz2016structural}, a machine-learned predictor of the propensity of a particle to rearrange. Softness is a weighted sum of quantities that characterize local structure that has provided useful insight into the mechanical response of disordered solids, including the size of particle rearrangements, origin of a universal yield strain~\cite{cubuk2018Science}, and precursors to shear banding~\cite{ivancic2019identifying, yang2020effect}.

Zhang {\it et al.}~\cite{zhang2021interplay} established a simple equation describing the average change in softness due to rearrangements. Here we incorporate softness and that equation, as well as, the connection between softness and local yield strain, to develop a ``structuro- elasto-plastic" (StEP) model for the response of the system to applied strain.
Our model connects the microscopic observations of Ref.~\cite{zhang2021interplay} to the collective response with no tunable parameters, in contrast to EP models that typically contain parameters fitted to reproduce collective response observed in particle simulations \cite{nicolas2017deformation}.  We obtain good quantitative agreement with simulations, validating the model. 

\begin{figure}
\includegraphics[width=0.5\textwidth]{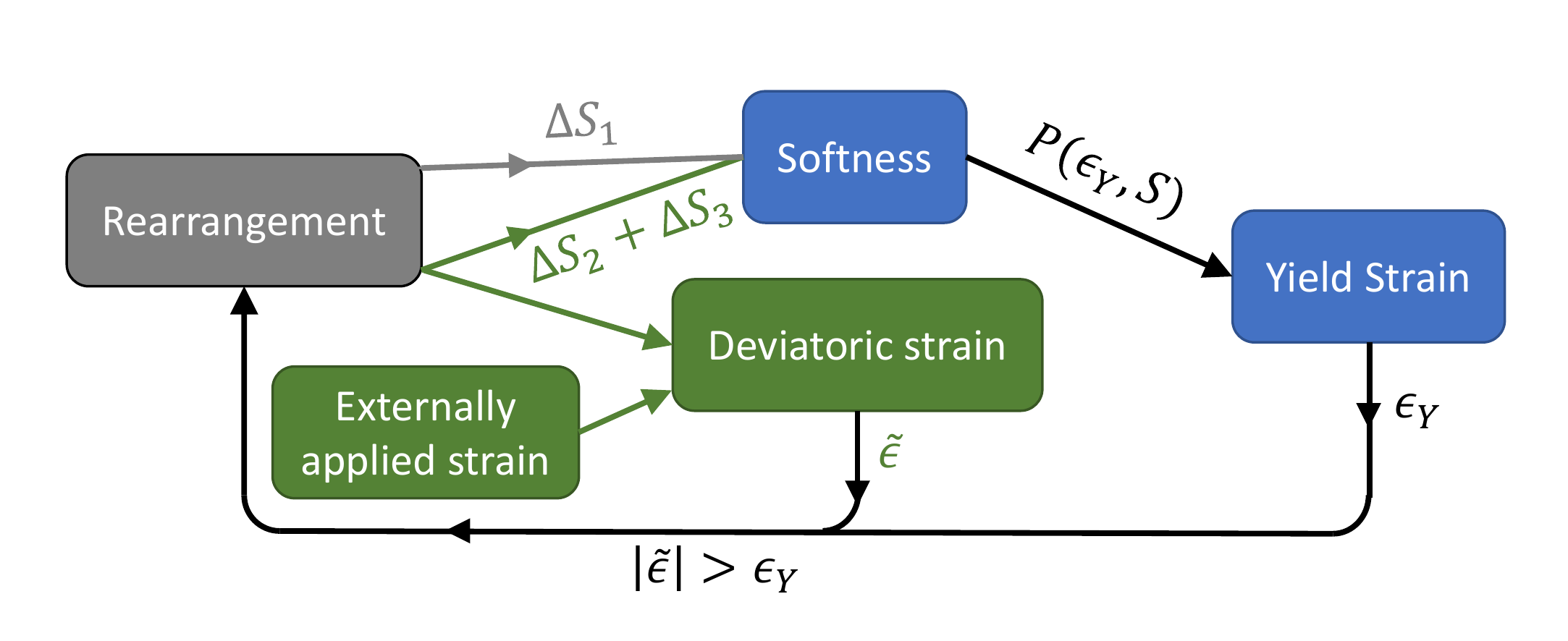}
\caption{Schematic of our model. A strain release (plastic event) at a given block changes the softness of nearby blocks, and propagates a deviatoric strain field to all other blocks. Softness determines a yield strain distribution, which in turn determines the yield strain of a block. A new rearrangement is triggered if the deviatoric strain is larger than the yield strain. Structural components are in blue, elasticity components are in green, and plasticity components are in gray; each arrow represents an equation.
}
\label{fig:flowChart}
\end{figure}

{\bf Particle Simulations and StEP Model:}
The ``ground truth" for this study is provided by simulations~\cite{zhang2021interplay} of two-dimensional systems of particles with short-ranged pairwise Hertzian repulsions: $U(r)= (1-r/\sigma)^{5/2}$, where $\sigma$ is the sum of the radii of the two interacting particles. We choose length and energy units such that the small particle diameter and the maximum interaction between two particles are unity. Each configuration contains $N_A=50,000$ small particles and $N_B=50,000$ large particles with radius ratio $r_B/r_A=1.4$. We repeatedly apply a small strain step of $\delta \epsilon=\ee{-5}$ and minimize the potential energy to approximate quasistatic athermal shear. Details of how we identify rearrangements and obtain softness are provided in the Supplementary Material~\cite{supplement}.

We calculate local yield strain by shearing the system in a certain orientation $\theta$, while only allowing neighbors within $R_c=2$ to move freely, and neglecting far-away particles that are not interacting with any particles inside $R_c$~\cite{barbot2018local}. The first point at which the local stress-strain curve $\sigma_\theta(\epsilon)$ decreases marks the local yield strain in this direction, $\epsilon_{Y, \theta}$. The particle's overall local yield strain, $\epsilon_Y$ is the minimum of $\epsilon_{Y, \theta}$ over $\theta$ of $\epsilon_{Y, \theta}$. In simulations we used $\theta=\{-85^\circ, -75^\circ, \ldots, 85^\circ\}$.

\begin{figure}
\includegraphics[width=0.49\textwidth]{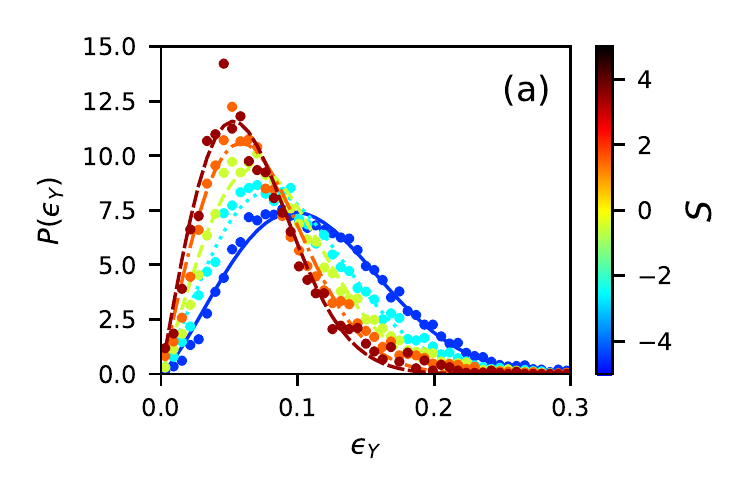}

\vspace{-0.2in}

\includegraphics[width=0.49\textwidth, trim=0 5 0 5, clip]{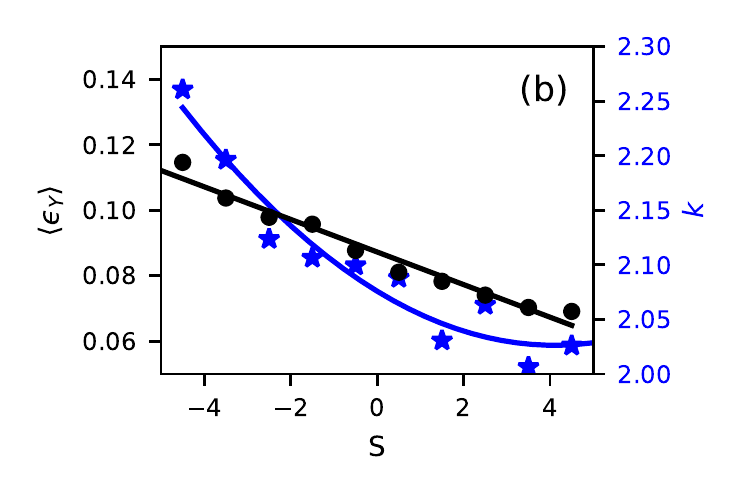}
\caption{(a) Probability density function (PDF) of local yield strain for particles with various softness values. Curves are Weibull distribution fits. (b) Mean local yield strain (black dots) and shape parameters (blue stars) for each softness bin. Curves are lowest-order polynomial fits, giving $\langle \epsilon_Y \rangle=0.087-0.0050S$ and $k=2.08-0.024S+0.0029S^2$, respectively.
}
\label{fig:yieldStrainStatistics}
\end{figure}

The distributions of local yield strain for particles with different softness values are plotted in Fig.~\ref{fig:yieldStrainStatistics}a. We see that softer particles tend to have lower local yield strain values, as expected. For each softness, the distribution is well described by the Weibull distribution~\cite{karmakar2010}:
\begin{eqnarray}
P(\epsilon_Y,S)=\frac{k}{\lambda}\left(\frac{\epsilon_Y}{\lambda}\right)^{k-1}\exp\left[-(\epsilon_Y/\lambda)^{k}\right]
\label{eq:weibull}
\end{eqnarray}
where $k(S)$ and $\lambda(S)$ characterize the distribution at each softness $S$. Equivalently, one can characterize the Weibull distribution in terms of $k$ and the mean local yield strain, $\langle \epsilon_Y \rangle=\lambda \Gamma(1+1/k)$, where $\Gamma(x)$ is the gamma function.
Fig.~\ref{fig:yieldStrainStatistics}(b) shows the shape parameter $k(S)$ and mean local yield strain $\langle \epsilon_Y \rangle(S)$. These functions relate softness to local yield strain.

The StEP model is summarized in Fig.~\ref{fig:flowChart}. The system is a square lattice of blocks. Each block is characterized by the local deviatoric strain (a 2D vector, $\tilde \epsilon$), the softness (a scalar, $S$), and a yield strain percentile (a scalar, which we will detail later). When the system is strained, the $xy$-strain is increased for each block with each strain step. 
\emph{Beyond this, our model deviates from typical EP models in several ways}. First, the local yield strain is no longer an explicit variable with a distribution that is immutable and crafted by hand. Rather, it is related to softness. We initially assign a softness, $S_i$, which determines the distribution of local yield strains to draw from. Upon randomly selecting $\epsilon_Y(i)$ from its distribution, we also store the percentile of the yield strain distribution corresponding to the sampled $\epsilon_Y(i)$. When a block rearranges, its yield strain percentile is re-drawn and stored; otherwise it is fixed. 

Second, we explicitly take into account both components of the deviatoric strain, not simply the $xy-$strain. This is because Ref.~\cite{zhang2021interplay} found that rearrangements are triggered by the deviatoric strain, not just the $xy$-strain, arising from other rearrangements. Once the deviatoric strain exceeds $\epsilon_Y(i)$ and the block rearranges, it loses all of its elastic strain while the local deviatoric strain of each other block is updated through the elastic kernel function calculated using the ``Fourier discretized'' method \cite{budrikis2013avalanche} 


Third, we set the block side length equal to the small particle diameter, rather than some mesoscopic scale. This is because softness is a particle-based quantity. While softness can be coarse-grained, and it is known that tuning block size is important for obtaining good quantitative agreement with simulations in a standard EP model~\cite{martens2021elastoplastic, castellanos2021insights, castellanos2022history}, we choose to fix it based on physical grounds instead of treating it as a parameter that can be tuned to improve agreement with particle-based simulations. Our system size is $L=360$ blocks per side, consistent with the side length of the particle simulations.

One result of this choice is that we cannot assume a rearrangement, which inevitably involves more than one particle, resides in a single block. 
As detailed in the Supplementary Material~\cite{supplement}, we find that the initial decay of the instant-time $\dtwomin$ correlation function in the particle simulations is well fit by $C(r)=\exp(-r/\xi)$ where $\xi=1.40$ 
provides a measure of the rearrangement size. To incorporate these correlations into the StEP model, we assume that when a block rearranges, other blocks within a distance $r<10$ also release strain with probability $C(r)$.
These strain releases are part of the rearrangement at $r=0$, rather than multiple independent rearrangements.

Fourth, to facilitate quantitative comparison with the simulations, we fix the timestep in the StEP model to match the simulations. As a consequence, we cannot assume that a plastic event occurs in a single timestep. As detailed in the Supplementary Material~\cite{supplement} the distance between two time frames in the simulations corresponds to a strain release of about $0.1$ per unit volume.  
To match this, in the StEP simulation we sequentially choose the site with the largest stress overshoot, $\abs{\tilde \epsilon}^2-\epsilon_Y^2$, and let it rearrange over a period of $t=\ceil{\sqrt{\sum \abs{\tilde \epsilon}^2}/0.1}$ time steps 
where $\ceil{x}$ represents the smallest integer that is larger than $x$, the sum is over all sites that release strain, and $\epsilon$ is the strain released.
The amount of strain released in each timestep is thus ${\tilde \epsilon}/t$.

Fifth, and most important, the model allows the softness field to evolve under strain. From previous simulations~\cite{zhang2021interplay} we know that the softness field changes whenever a rearrangement occurs, with two contributions, $\Delta S=\Delta S_1 + \Delta S_2$, corresponding to the near-field plastic and far-field elastic responses to a rearrangement at the origin, respectively. The first term describes the near-field plastic effect of a rearrangement on softness~\cite{zhang2021interplay} and has the form $\Delta S_{1}=-0.3r^{-3.1}+0.06r^{-3.2}(\langle S \rangle-S)$. The two exponents in this equation, $-3.1$ and $-3.2$, were obtained from numerical fits. We assume their small difference is not significant and simplify this expression to
\begin{equation}
 \Delta S_1(r)=\eta(r)(\langle S \rangle-S+c),
\label{eq:s1}   
\end{equation}
where $\eta(r)=0.06r^{-3.1}$ and $c=-5$. 

The second term is proportional to the far-field volumetric strain, $k(\mathbf r)$, caused by the rearrangement. Since the primary effect of this volumetric strain is to change softness, it suffices to track only that change of softness and not the volumetric strain itself. Ref.~\cite{zhang2021interplay} showed that for the system studied here, this effect is well described by $\Delta S_2\approx 207k(r)$ where $k(r)=(\nu-1)\abs{\tilde \epsilon}\sin(2\theta)/2\pi r^2$, $\nu=0.443$ is the Poisson ratio and $\theta$ is the orientation of $\mathbf r$ relative to the orientation of the strain release. 
Combining these two equations yields
\begin{eqnarray}
\Delta S_2(r)=a\abs{\tilde \epsilon}\sin(2\theta)r^{-d},
\label{eq:s2}   
\end{eqnarray}
where $a=-18.3$ and $d=2$ is the spatial dimension, for a strain release of $\abs{\tilde \epsilon}$. 

Eqs.~(\ref{eq:s1},\ref{eq:s2}) are empirical formulae obtained from particle simulations for the softness change in response to a rearrangement at the origin, but require three adjustments to be implemented in the StEP model. 
The first adjustment is to the interpretation of $\langle S \rangle$ in Eq.~\ref{eq:s1}. It is unreasonable to assume that $\Delta S_1$ is affected by the softness field very far away in a large system. This problem is particularly acute in systems without time-translation symmetry, in which $\langle S \rangle$ changes with time. To fix this problem, we replace $\langle S \rangle$ with $\langle S(r) \rangle$, the average of softness of particles at distance $r$ to a rearranger over the most recent 1000 timesteps.

To understand the second adjustment, note that Zhang {\it et al.}'s analysis~\cite{zhang2021interplay} was applied to avalanches (stress drops) in quasistatically sheared systems, where no strain is applied during the avalanche. It neglects the effect of elastic deformation on the softness field, which is known to increase softness~\cite{cubuk2018Science}. We have measured the elastic part of the stress-strain relation in our particle simulations (the initial linear part of the stress-strain curve) and find that this ``loading" contribution can be summarised as the first term below:
\begin{equation}
   \Delta S_3=b\Delta(\tilde \epsilon^2) + c^\prime,
   \label{eq:s3}  
\end{equation}
where $b=101.3$ and $\Delta(\tilde \epsilon^2)$ is the change of square of the elastic deviatoric strain, by symmetry. To understand the second term in Eq.~\ref{eq:s3}, note that on average, Eq.~(\ref{eq:s1}) causes the average softness of the whole system to drop, but this is not observed in particle simulations.  This suggests that there exists a term that counters the drop, which is likely spread over a sufficiently large area that it is too small to be observed in the particle simulations in \cite{zhang2021interplay}. To represent this term, we add a small constant softness change $c'=-c\langle \eta(r) \rangle$ to all sites, where $\langle \eta(r) \rangle$ is $\eta(r)$ numerically averaged over all sites.


\begin{figure*}
\includegraphics[width=0.32\textwidth, trim=10 0 10 0, clip]{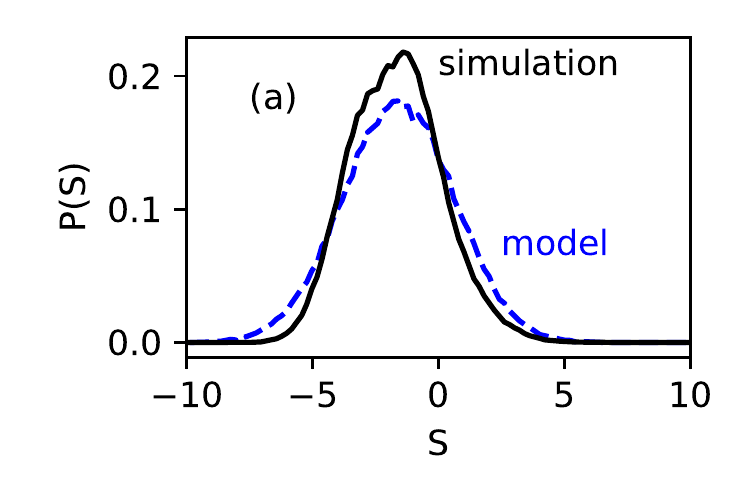}
\includegraphics[width=0.32\textwidth, trim=10 0 10 0, clip]{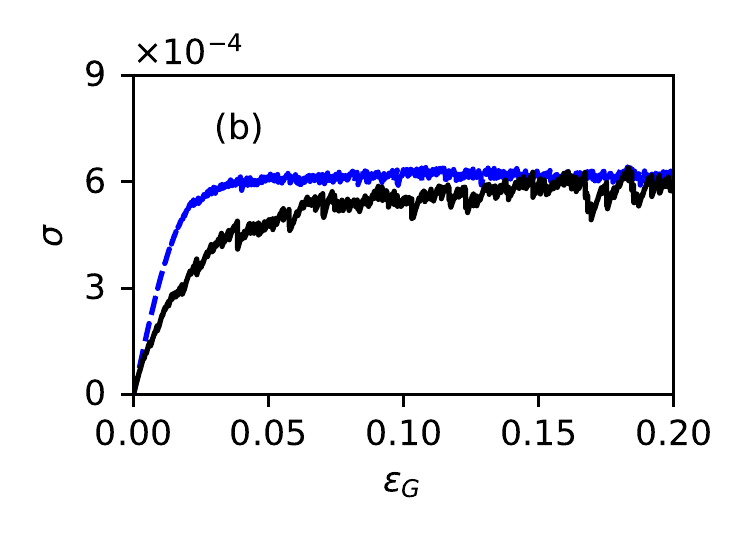}
\includegraphics[width=0.32\textwidth, trim=12 0 10 0, clip]{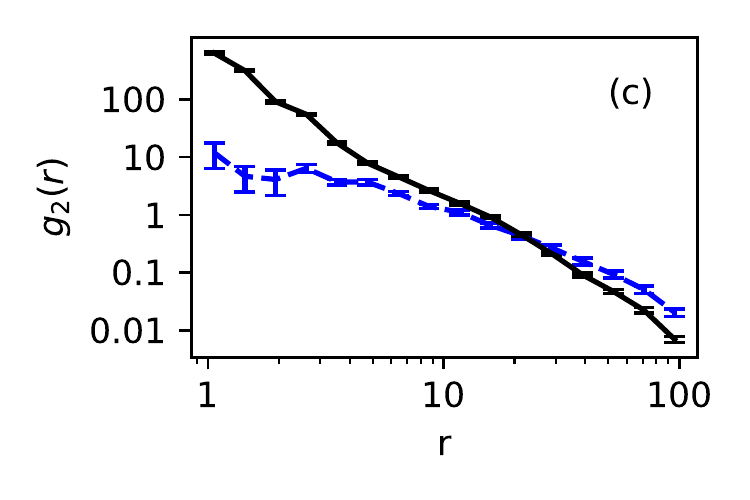}
\caption{Statistics of the StEP model (blue dashed line) compared with particle simulations (black solid line), including the (a) softness distribution, (b) stress-strain curve, 
and (c) pair correlation function of rearrangers (right). 
}

\label{fig:avalancheStatistics}
\end{figure*}

Finally, Eqs.~(\ref{eq:s1},\ref{eq:s2}) describe the {\it average} softness change. To find the softness of a given particle (block), we have added a Gaussian-distributed random noise term, $\delta(r)$. Its magnitude is derived using a detailed balance argument~\cite{supplement}. That argument normally applies to a system in thermal equilibrium, but even for quasistatically sheared systems, an effective temperature well describes particle-level dynamics \cite{ono, berthier2002shearing, song2005experimental, haxton}.  

In summary, the softness change of a particle with softness $S$, at distance $r$ from a rearrangement releasing strain $\epsilon$ is
\begin{equation}
\begin{split}
    \Delta S(r, S, \epsilon)&=\eta(r)(\langle S(r) \rangle-S+c)\\&+a\epsilon\sin(2\theta)r^{-\lambda}+b\Delta(\tilde \epsilon^2)\\&+c'+\delta(r)
\end{split}
\label{eq:totalDeltaS}
\end{equation}
where all terms are established from particle simulations.

To put the StEP model in the context of previous models as reviewed in Table I of Ref.~\cite{nicolas2017deformation}, we include nontrivial barrier (yield strain) distributions, our plastic events have a nonzero time duration and we use elastic propagators.

{\bf Results:}
The process of an avalanche in our StEP model, as well as the pattern of all rearrangements during an entire avalanche, is similar to that in particle simulations \cite{zhang2021interplay}. At any given time, there are only a few localized rearranging blocks, but the total avalanche can be extended as one event triggers another~\cite{supplement}. 

We now compare quantitative results from the StEP model with the particle simulations (Fig.~\ref{fig:avalancheStatistics}). The steady-state softness distribution is an emergent property that does not depend on the initial distribution. Here we calculate $P(S)$ 
by averaging over the entire run in both the StEP model and the simulations (Fig.~\ref{fig:avalancheStatistics}a). The softness distribution from the StEP model (blue dashed curve) is in good quantitative agreement with that of the particle simulations (black solid curve).

The stress-strain curves are likewise in good agreement, as are the pair correlation function of rearrangers (Fig.~\ref{fig:avalancheStatistics}b-c). Note that while the stress-strain curve rises more sharply for the StEP model, it is monotonic with fluctuations, indicating that the system is ductile, in agreement with the simulations. The pair correlation function of rearrangers compares results over shorter length scales. Its agreement for $r \gtrsim 5$ demonstrates the success of our model over those length scales. In contradistinction, standard EP models are generally considered mesoscopic, with block side lengths larger than five times particle diameters \cite{nicolas2017deformation}.

We also calculate the scaling exponents of the avalanche size distributions and compare to particle simulations for 2D overdamped Lennard-Jones systems\cite{salerno2012avalanches}, since the exponents should not be sensitive to interaction potential. For particle simulations, the rate of an energy drop of size $E$ follows the finite-size scaling ansatz $R(E, L)=L^\beta g(E/L^\alpha)$, where $L$ is the system length, $g(x)$ is an unknown function that scales as $x^{-\tau}$ for small $x$, and $\alpha$ and $\beta$ are exponents satisfying $\beta+2\alpha=d$ in $d$ dimensions \cite{salerno2012avalanches}. Defining $\gamma=\beta+\alpha\tau$, then $R(E, L)\propto L^\gamma E^{-\tau}$ for small $E$. Salerno and Robbins~\cite{salerno2012avalanches} found $\alpha=0.9\pm0.05$, $\beta=0.2\pm0.1$, $\gamma=1.3\pm0.05$, and $\tau=1.25\pm0.05$. 
As shown in Fig.~S3 in the Supplementary Material~\cite{supplement},
excellent agreement with the ansatz can be found for the StEP model with exponents $\alpha=0.98$, $\beta=0.04$, $\gamma=1.30$, and $\tau=1.29$. The ansatz holds for not only the energy-drop rate, but also for the rate of the rescaled stress drops $\Sigma=\sigma L^d$. Using $\Sigma$, we obtain scaling exponents $\alpha=0.99$, $\beta=0.02$, $\gamma=1.29$, and $\tau=1.28$. Clearly, the exponents $\alpha$, $\gamma$ and $\tau$ are very similar for the StEP and particle-based calculations. 

{\bf Discussion:}
In this paper, we have explicitly incorporated local structure into a standard elastoplastic (EP) framework to build a structro-elasto-plasticity (StEP) model. Elastoplastic models typically contain the effects of long-ranged elastic facilitation triggered by elastic strain arising from a rearrangement. Our model contains not only long-ranged facilitation
but also short-ranged facilitation due to near-field plastic effects. The effects of such short-ranged facilitation have recently been characterized in detail~\cite{richard2022}. Both facilitation terms are characterized fully by parameters directly measured from numerical simulations, and produce results quantitatively similar to those of the simulations without any adjustment. 
The inclusion of structural information into EP models is an important natural extension of such models, and the quantitative agreement we have achieved suggests that our model captures the key physics. Our model can be used to gain microscopic understanding of the factors that control nonlinear mechanical response. Changes in preparation history, for example, should change only the softness distribution, while other changes will be reflected in Eq.~\ref{eq:totalDeltaS}, which quantifies short-ranged and elastic facilitation.  The next step is to build StEP models for particle-based systems that can be tuned to exhibit behaviors over the spectrum of ductile to brittle response.

We thank R. C. Dennis and M. L. Manning for helpful discussions. This work was supported by the National Science Foundation through grant MRSEC/DMR-1720530 (DJD, RAR, HX, EY, GZ), and the Simons Foundation via the ``Cracking the glass problem" collaboration \#45945 (SAR, AJL) and Investigator Award \#327939 (AJL).

\bibliography{references}

\end{document}